\documentclass{article}
\usepackage{spconf,amsmath,graphicx}
\usepackage{pifont,subfigure,enumitem,cite}
\usepackage{amssymb}

\title{Incremental Learning for End-to-End Automatic Speech Recognition}
\name{Li Fu, Xiaoxiao Li, Libo Zi, Zhengchen Zhang, Youzheng Wu, Xiaodong He, Bowen Zhou}
\address{JD AI Research, Beijing, China}

\begin{document}
%
\maketitle
\begin{abstract}
In this paper, we propose an incremental learning method for end-to-end Automatic Speech Recognition (ASR) which enables an ASR system to perform well on new tasks while maintaining the performance on its originally learned ones. To mitigate catastrophic forgetting during incremental learning, we design a novel explainability-based knowledge distillation for ASR models, which is combined with a response-based knowledge distillation to maintain the original model's predictions and the ``reason'' for the predictions. Our method works without access to the training data of original tasks, which addresses the cases where the previous data is no longer available or joint training is costly. Results on a multi-stage sequential training task show that our method outperforms existing ones in mitigating forgetting. Furthermore, in two practical scenarios, compared to the target-reference joint training method, the performance drop of our method is 0.02$\%$ Character Error Rate (CER), which is 97$\%$ smaller than the drops of the baseline methods.
\end{abstract}
\section{Introduction}
\label{sec:intro}

While end-to-end Automatic Speech Recognition (ASR) models have been widely used in many applications \cite{ref1,ref2}, they usually suffer from performance degradation when applied to new tasks with new accents, new words, or new acoustic environments, etc. Thus, it is worth adapting a pre-trained ASR model to new tasks while maintaining its performance on original tasks. One real-world case is to enable an ASR model trained on native speakers to also recognize speeches with foreign accents. Another case is to add the most recent internet slang to a legacy model’s recognition ability. Note that for both scenarios, any modifications to the pre-trained ASR models should not degrade their performance on the previously learned tasks.

Naively, one can jointly train the acoustic model on the union of the new task data and the original task data. However, with the continually emergence of new tasks for ASR, joint training will be required repeatedly which is practically infeasible due to the usually high training cost. In addition, the original task data may be unavailable due to security and privacy issues \cite{ref5}. An alternative solution is to apply a specifically designed language model for decoding schemes or post corrections; but this approach is limited by some essential problems arising from the acoustic model which are still unsolved \cite{ref6}. Fine-tuning is usually used to adapt a pre-trained model to new tasks by modifying its parameters solely based on training data from new tasks \cite{ref7}. However, the performance of a fine-tuned model on original tasks may be degraded due to the absence of mechanisms preventing the model from forgetting. 

Incremental learning studies the problem of gradually learning on new tasks while maintaining the existing knowledge \cite{ref8,ref9}. Based on incremental learning, we propose a novel training method for end-to-end ASR models in adaptation to new tasks without obvious forgetting. In particular, our method only uses the pre-trained model and the data from new tasks, but {\it does not} use any data from original tasks, which reduces training cost and addresses data privacy issues. The advantages of our method over directly using the pre-trained model, joint training, and fine-tuning are summarized in Table \ref{table1}.

To maintain a model's previous knowledge without access to any data from the original task, one can align the model's {\it behavior} with the model pre-trained on the original task. Thus, we adopt a teacher-student framework \cite{ref10}, where the teacher model is the pre-trained ASR model; while the student model is initialized with the pre-trained ASR model. Ideally, one can force the outputs of the two models to be similar for a sufficiently abundant input set with high modality. However, in our incremental learning scenario, alignment of the two models can only exploit the data from the new task, which has limited similarity with the data from the original task in terms of data distribution. Empirically, such similarity can be exploited by Response-based Knowledge Distillation (RBKD) mechanisms to help the student model align the predictions of the teacher model on original tasks \cite{ref3}. However, due to the difference in distribution between the new task data and the original task data, we believe that only aligning the predictions of the teacher model and the student model using RBKD is not sufficient for the student model to adequately learn the teacher model's behaviour \cite{ref21}. Thus, we propose a novel Explainability-based Knowledge Distillation (EBKD) to help the student model also learn the ``reason'' for the predictions produced by the teacher model. 

Specifically, to train the student model, we propose a novel training loss function involving the following three terms: 1) a term associated with a conventional Connectionist Temporal Classification (CTC) loss \cite{ref11}, which is adopted to help the ASR model learn on the new task; 2) a term associated with a novel EBKD proposed for the ASR incremental learning; 3) a term associated with a RBKD from the teacher model. In particular, the EBKD term and the RBKD term are designed to maintain the pre-trained model’s predictions and the “reason” for the predictions. Experimentally, for ASR incremental learning tasks, the novel EBKD term significantly improves the performance of our method in comparison with the baseline methods involving only the RBKD term \cite{ref12}. Our main contributions are shown as follows:

1. We propose a new incremental learning method for end-to-end ASR to improve the model’s performance on new tasks while maintaining its previous knowledge. Compared with the joint training method, ours does not require the training set of the pre-trained model, thus is computationally more efficient.

2. We propose a novel EBKD for ASR incremental learning which largely reduces the CERs compared with the baseline methods \cite{ref12} on a multi-stage sequential training task.

3. We also evaluate the effectiveness of our method in two practical scenarios: for the ASR model to incrementally learn a new accent and new words, respectively; our method outperforms the baseline methods \cite{ref12} with obvious CER decreasing.

\begin{table}[t]
	\centering  
	\caption{Advantages of Incremental Learning (IL) without access to Original Task (OT) data over: directly using the pre-trained model, joint training, and fine-tuning, in terms of access to OT data or New Task (NT) data, training cost, performance on OTs or NTs. Generally, IL requires no access to OT data, requires minimum training cost, while showing good performance on both OTs and NTs.}
	\label{table1}  
	\resizebox{0.95\linewidth}{!}{
	\begin{tabular}{c|c|c|c|c}  
		\hline  
		&\begin{tabular}{c}
		    \textbf{Pre-}\\
		    \textbf{training}
		\end{tabular}
		&\begin{tabular}{c}
		    \textbf{Joint}\\
		    \textbf{training}
	    \end{tabular}
		&\begin{tabular}{c}
		    \textbf{Fine-}\\
		    \textbf{tuning}
		\end{tabular}
		&\begin{tabular}{c}
		    \textbf{IL}
		\end{tabular} \\
		\hline
		\hline 
		Access to OT data & Yes & Yes & No & No \\
		\hline
		Access to NT data & No & Yes & Yes & Yes \\
		\hline
		Training cost & High & High & Low & Low \\
				\hline
		Performance on OTs & Good & Good & Poor & Good \\
				\hline
		Performance on NTs & Poor & Good & Good & Good \\
		\hline
	\end{tabular}
	}
\end{table}

\section{Related Work}
\label{sec:relatedwork}

Existing incremental learning methods mainly fall into three categories, based on how the original task data is used: a) original task data is involved in generating synthetic data for training; b) original task data is sampled when constructing a new data set for training; c) original task data is {\it not} used for training \cite{ref13}. Although methods from the first two categories may mitigate the forgetting by leveraging the knowledge of original task data, sophisticated algorithms for data synthesis and/or sampling are required. Moreover, these methods fail when the original task data is not available due to data privacy issues.

In this paper, we focus on the category of incremental learning methods {\it without access to} original task data. Existing methods usually adopt RBKD and/or Elastic Weight Consolidation (EWC) for maintaining the previous knowledge \cite{ref14}. In comparisons, for ASR domain extension tasks, RBKD usually outperforms EWC \cite{ref12}. One possible reason is that EWC applies overly tight constraints to the model parameters, which likely causes impaired learning for new tasks \cite{ref15}. While we focus on end-to-end ASR models, incremental learning methods for hybrid ASR models based on RBKD and/or EWC are also explored and have achieved clear improvements over fine-tuning \cite{ref16}. Also for hybrid ASR models, \cite{ref17} proposed to modify the model architecture by adding new parameters during training. However, increment of model complexity usually causes higher time consumption during inference.

\section{Incremental Learning for ASR}
\label{sec:ILASR}

\subsection{Problem description}
Given an ASR model pre-trained on original task data $D_1$ and a data set $D_2$ associated with a new task (different from $D_1$ in terms of accents, words, or acoustic environments, etc), the incremental learner aims to train a new ASR model that:

\begin{itemize}[leftmargin=0.12in]
	\setlength{\itemsep}{-2pt}
	\item performs well on the new task associated with $D_2$;
	\item has similar performance with the pre-trained ASR model on the original task associated with $D_1$.
\end{itemize}

Note that the learner has access to the pre-trained ASR model and $D_2$, but {\it has no access to} $D_1$ used for model pre-training. For convenience, we denote ${D}_{2}=\{\mathbf{x}^{i},\mathbf{y}^{i}|i\in \{1,\cdots,{N}\}\}$, with ${N}$ the number of samples; $\mathbf{x}^{i}\in\mathbf{R}^{{F}\times{S}_{i}}$ is the $i^{th}$ sample of ${D}_{2}$ which is a sequence of $F$-dimensional acoustic features with length ${S}_{i}$; $\mathbf{y}^{i}\in\mathbf{L}^{{U}_{i}}$ is the associated label sequence with length ${U}_{i}$, with $\mathbf{L}$ the finite label alphabet.


\begin{figure}[t]
  \centering
  \includegraphics[height=7cm]{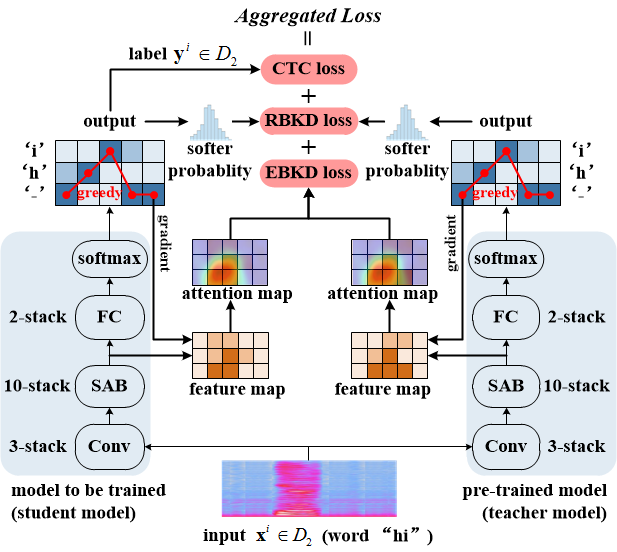}

  \caption{Outline of our incremental learning method for end-to-end ASR. We adopt a teacher-student framework, where the teacher model (right) is the pre-trained ASR model; while the student model (left) is initialized with the pre-trained ASR model. The training loss of our method is aggregated from three terms: a CTC loss solely based on the model to be trained (student model), which is introduced to help the model learn on the new task; two distilling losses (consisting of a RBKD loss and an EBKD loss) leveraging the pre-trained model (teacher model), which are designed to maintain the model's performance on the original task.}
  \label{fig1}
\end{figure}

\subsection{Incremental learning method}
We propose a novel loss function for incremental learning, which contains a CTC loss \cite{ref11}, a RBKD loss, and an EBKD loss. The CTC loss is introduced to help the model learn on the new task; while the RBKD loss and the EBKD loss are designed to maintain the model's performance on the original task. Note that we adopt CTC loss and associated ASR model architectures here only for an example. Our work can be easily extended to alternative end-to-end ASR model frameworks, such as Recurrent Neural Network Transducer (RNN-T) \cite{ref18}, Listen Attend and Spell (LAS) \cite{ref19}, etc. Similar to \cite{ref20}, we consider ASR models involving Self-Attention Blocks (SABs). For example, our experiments use a model architecture with a successive stack of 3 Convolutional (Conv) layers, 10 SABs, 2 Fully Connected (FC) layers, and a feature-wise softmax activation. In this section, we drop the model parameters for simplicity, which are described in detail in Section 4.1. The outline of our method is summarized in Figure \ref{fig1}.


{\bf CTC loss:} For any labeled sample $\{\mathbf{x}^{i},\mathbf{y}^{i}\}\in{D_2}$, denote the softmax output of the model as $[\boldsymbol{\pi }_{v}^{1},\cdots,\boldsymbol{\pi }_{v}^{k},\cdots,\boldsymbol{\pi }_{v}^{K}]$, where $K$ is the sequence length and $\boldsymbol{\pi}_{v}^{k}\in\mathbf{R}^{M}$ is the probability mass with $M$ the alphabet size plus one blank symbol; $v=1$ and $v=2$ denote the pre-trained model and the model to be trained, respectively. A valid CTC path with $K$ symbol sequence $c=(c_1,\cdots,c_K)$ is a variant of the true label $\mathbf{y}^{i}$ that allows occurrences of blank symbols and repetitions. Then, the CTC loss can be written as \cite{ref11}:

\begin{equation}
  {{L}_{\rm CTC}}=-{\rm log}(\sum_{c\in {\mathcal C(\mathbf{x}^{i},\mathbf{y}^{i})}}\prod_{k=1}^{K}\boldsymbol{\pi }_{2,c_k}^{k})
  \label{eq1}
\end{equation}
where $\mathcal C(\mathbf{x}^{i},\mathbf{y}^{i})$ is the set of all valid CTC paths; $\boldsymbol{\pi }_{v,c_k}^{k}$ is the probability corresponding to the symbol $c_k$ in the vector $\boldsymbol{\pi }_{v}^{k}$.





\begin{figure*}[t]
	\centering
	\subfigure[Parameter $T$ ($\beta=0.03,\gamma \text{ = 500}$)]{
		\begin{minipage}[t]{0.3\linewidth}
			\centering
			\includegraphics[width=5.3cm]{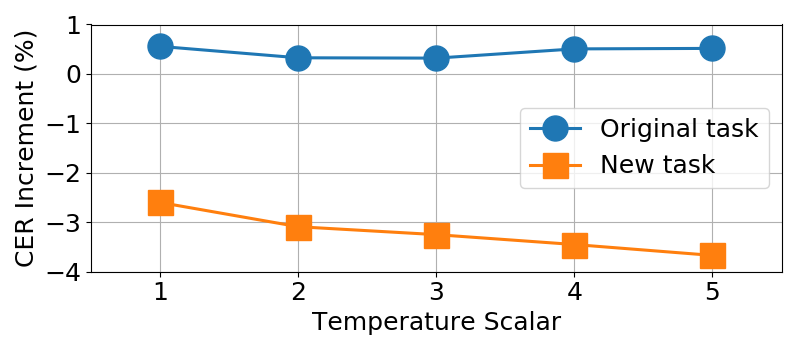}
		\end{minipage}
	}
	~
	\subfigure[Parameter $\beta$ ($T=3,\gamma \text{ = 500}$)]{
		\begin{minipage}[t]{0.3\linewidth}
			\centering
			\includegraphics[width=5.3cm]{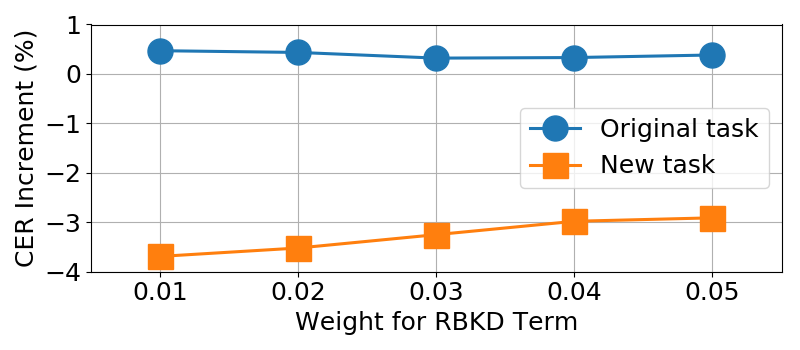}
		\end{minipage}
	}
	~
	\subfigure[Parameter $\gamma$ ($T=3,\beta =0.03$)]{
		\begin{minipage}[t]{0.3\linewidth}
			\centering
			\includegraphics[width=5.3cm]{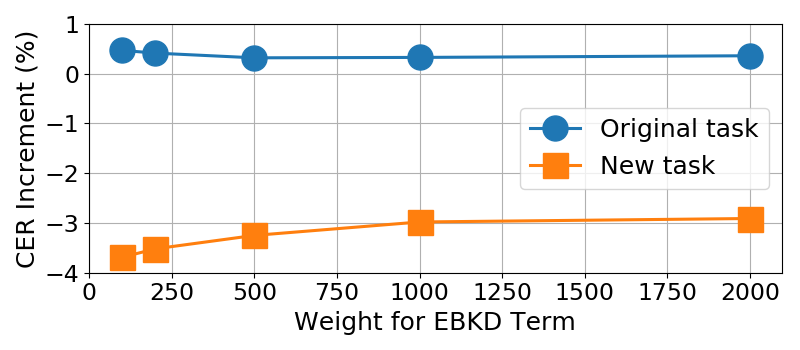}
		\end{minipage}
	}%
	\centering

	\caption{Compared to the pre-trained model, the CER increments of the model trained by our incremental learning method on both original tasks and new tasks (Performance improvements are indicated by negative CER increments), for a range of choices of (a) the temperature scalar $T$; (b) the weight $\beta$ for the RBKD term; and (c) the weight $\gamma$ for the EBKD term.}
	\label{fig2}
\end{figure*}

{\bf RBKD loss and EBKD loss:} To maintain a model's previous knowledge without access to any data from the original task, one can align the model's {\it behavior} with the model pre-trained on the original task. As discussed in Section 1, we employ a RBKD and an EBKD mechanism to help the student model to learn the prediction and the "reason" behind the prediction of the teacher model. 

To guide the student model to produce similar predictions as the teacher model, we use the same RBKD loss as in \cite{ref12}:

\begin{equation}
  {{L}_{\rm RBKD}}=-\sum\limits_{k=1}^{K}{\sum\limits_{m=1}^{M}{ {p}_{1,m}^{k}\log ({p}_{2,m}^{k})}}
  \label{eq2}
\end{equation}
where ${p}_{v,m}^{k}= \boldsymbol{(\pi }_{v,m}^{k})^{1/T}/\sum\limits_{m=1}^{M}{\boldsymbol{(\pi }_{v,m}^{k})^{1/T}}$, and $T$ is a temperature scalar that produces the softer probability.

To help the student model also learn the ``reason'' for the predictions produced by the teacher model, 
we propose a novel EBKD loss for ASR incremental learning to train the student model to have similar attention maps \cite{ref22} to those of an already trained teacher model. Specifically, for input $\mathbf{x}^{i}$, we choose model $v$'s output of the last SAB layer as the feature map, which is denoted as $\mathbf{A}_{v}^{{}}\in\mathbf{R}^{d_h\times K}$ with hidden dimension $d_h$. Then we calculate the importance of each element in $\mathbf{A}_{v}^{{}}$ to the model's most-probable prediction by taking the gradient of the model's greedy prediction probability $p_{v}=\underset{k=1}{\overset{K}{\mathop{\text{ }\!\!\Pi\!\!\text{ }}}}\,\underset{m}{\mathop{\max }}\,(\boldsymbol{\pi }_{v,m}^{k})$ over $\mathbf{A}_{v}^{{}}$, yields the importance map ${\boldsymbol{\alpha }_{v}}\in\mathbf{R}^{d_h\times K}$ \cite{ref22}:
\begin{equation}
  {\boldsymbol{\alpha }_{v}}=\partial \log (p_v)/\partial \mathbf{A}_{v}^{{}}
  \label{eq3}
\end{equation}

Note that one can also consider other choices of the feature map, or even a combination of multiple feature maps. Then, the attention map $\boldsymbol Q_{v}^{{}}\in\mathbf{R}^{d_h\times K}$ is defined as

\begin{equation}
  \boldsymbol Q_{v}^{{}}={{f}_{\operatorname{\rm ReLU}}}({\boldsymbol {\alpha }_{v}}\odot\mathbf{A}_{v}^{{}})
  \label{eq4}
\end{equation}
where ${{f}_{\rm {ReLU}}}(\cdot)$ is the element-wise ReLU function, which introduces a positive influence on the result of interest \cite{ref22}; and $\odot$ denotes element-wise multiplication. In our method, the attention map provides a reasoning associated with the high-level features of interest for the model's most-probable prediction, which is proved to be useful for the student model learning the behaviour of the teacher model. Finally, the attention map is normalized feature-wise which yields our EBKD loss:

\begin{equation}
  {{L}_{\rm EBKD}}=\frac{1}{K}\sum\limits_{k=1}^{K}{{{\left\| \frac{\boldsymbol Q_{2}^{k}}{{{\left\|\boldsymbol Q_{2}^{k} \right\|}_{2}}}-\frac{\boldsymbol Q_{1}^{k}}{{{\left\| \boldsymbol Q_{1}^{k} \right\|}_{2}}} \right\|}_{2}}}
  \label{eq5}
\end{equation}
where $\boldsymbol Q_{v}^{k}\in\mathbf{R}^{d_h}$ is the ${{k}^{th}}$ vector of $\boldsymbol Q_{v}^{{}}$.

{\bf Aggregated loss:} Loss for our incremental learning for end-to-end ASR is:
\begin{equation}
  L={{L}_{\rm CTC}}+\beta {{L}_{\rm RBKD}}+\gamma {{L}_{\rm EBKD}}
  \label{eq6}
\end{equation}
where hyperparameters $\beta $ and $\gamma $ are used to balance these terms -- their choices are experimentally studied in Section 4.3 (see Figure \ref{fig2}).

{\bf Deeper insight of the proposed EBKD loss:} For our incremental learning framework, there is no guarantee that the samples $\mathbf{x}^{i}$ for the new task follow exactly the same distribution as those for the original task. But the attention map generated by the teacher model for any sample $\mathbf{x}^{i}$ highlights the important internal layer features that are decisive to its output. Thus, our EBKD term, by its formulation, is designed to guide the student model paying more attention to these features ``reasoned'' by the teacher model.
With the EBKD term, the student model is then trained to give the same responses to these important features. Formally, $\boldsymbol Q_{1}^{{}}$ should be similar to $\boldsymbol Q_{2}^{{}}$.

\begin{figure}[t]
	\centering
	\subfigure[Distributions of EBKD loss on original task]{
		\begin{minipage}[t]{0.57\linewidth}
			\centering
			\includegraphics[width=\linewidth]{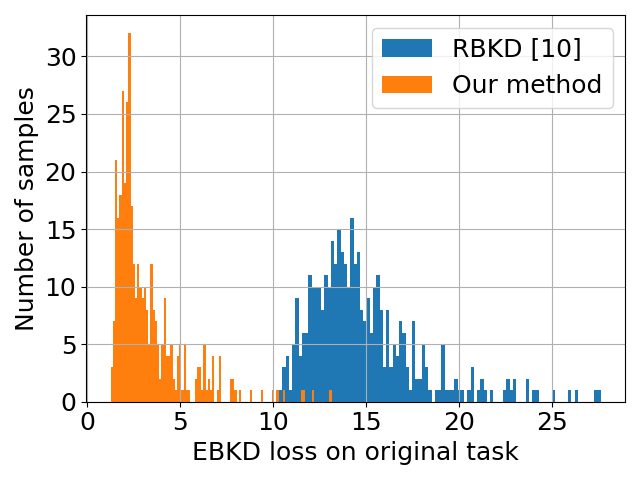}
		\end{minipage}
	}%
	~
	
	\subfigure[Relations between CER of each sample and EBKD loss on original task]{
		\begin{minipage}[t]{0.57\linewidth}
			\centering
			\includegraphics[width=\linewidth]{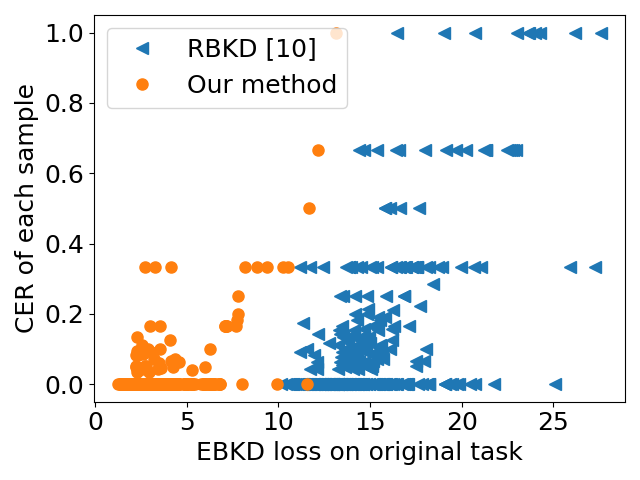}
		\end{minipage}
	}%
	\centering
	\caption{EBKD losses on original task (400 samples randomly selected) of our method and RBKD [10]: (a) shows the EBKD loss of our method decreases with convergence; (b) shows the correlations between CER of each sample and EBKD loss on the original task are 0.77 for these samples with recognition errors (CER$>$0) and 0.63 for all selected samples.}
	\label{fig4}
\end{figure}

Although visualization of the above mentioned effect for speech features is much more difficult than for the image domain \cite{ref22}, we find that there is a significant correlation between knowledge forgetting and EBKD loss in our incremental learning tasks. As shown in Figure \ref{fig4} (as an example derived from the main results in Table \ref{table4}), the recognition error on the original task and the value of the proposed EBKD loss show a strong correlation: 0.77 on average for our method and the baseline method involving only the RBKD term \cite{ref12}. 
Compared with the baseline method, the knowledge forgetting (CER increment on the original task) of our method is mitigated as our EBKD loss decreases with convergence -- {\bf the EBKD loss functions as expected.}

\section{Results and Discussion}
\label{sec:results}

\subsection{Experimental setup}

{\textbf{Data preparation:}} To evaluate our proposed method more thoroughly, we use a combinatory data set containing 12,301 hours of Mandarin speech from a variety of data subsets with different accents, contents/topics of the text, and total duration \cite{ref23,ref24,ref25,ref26,ref27}. Details for these data subsets are shown in Table \ref{table2}. For the public data, we use their original train-test split. For the other data subsets, we hold out 10\% speech samples for testing. Each speech is pre-processed as a sequence of 80-dimensional filter banks, with each element of the sequence obtained from a 20ms window (with 10ms shift) of the speech \cite{ref28}.

\textbf{ASR Model:} The model considered here contains 3 Convs, 10 SABs, and 2 FCs in sequential order (see Figure \ref{fig1}). For the three Convs, the filter sizes are (41,11), (21,11), and (21,11), respectively; the channel counts are 32, 32, and 96, respectively; and the strides are (2,2), (2,1), and (2,1), respectively \cite{ref11}. All 10 SABs have 8 multi-heads and are 256-dimensional. The output dimensions of the penultimate FC and the last FC are 1024 and 7229, respectively, where 7229 is the number of Chinese characters ever occurring in the entire data set plus a symbol blank. Following the convention of CTC-based ASR, a 4-gram language model, which is trained using the KenLM toolkit \cite{ref29}, is used for inference. The pre-training of the ASR model uses the CTC loss as in (\ref{eq1}) \cite{ref11}. Both the pre-training and our incremental training are trained with a mini batch of 128 and using the same learning rate scheduler as in \cite{ref20}.

\begin{table}[t]
	\centering  
	\caption{Details for each composition of the combinatory Mandarin speech data set used in our experiments.}  
	\label{table2}  
	\resizebox{1\linewidth}{!}{
	\begin{tabular}{c|c|c}  
		\hline  
		\textbf{Composition name}&\textbf{Duration (h)}&\textbf{Description} \\  
		\hline
		\hline
		JD in-house data&10,000&\begin{tabular}{c}
		     Conversational telephony \\
		     speech
		\end{tabular}  \\
		\hline
		Public data*&2000&Standard Mandarin speech \\
		\hline
		New accents data&300&Accent telephony speech \\
		\hline
		     New words data
		&1&\begin{tabular}{c}
		     Various new words of \\
		     internet bank speech
		\end{tabular} \\
		\hline
		\multicolumn{3}{l}{\begin{tabular}{l}
		     \scriptsize
		     *The public data consists of AISHELL-1 (170h) \cite{ref23}, AISHELL-2-IOS (1000h) \cite{ref24},\\
		     \scriptsize
		     THCHS-30 (40h) \cite{ref25}, Primewords (100h) \cite{ref26}, ST-CMDS (100h) \cite{ref27}, etc.
		\end{tabular}}
	\end{tabular}
	}
\end{table}

\begin{table*}[t]
	\centering  
	\caption{CERs of our method compared with fine-tuning, RBKD, RBKD with EWC, and joint training, on the Original Tasks (OTs) ever visited and the current New Task (NT), for each stage of the 5-stage sequential training which visits data sets $B_1$, $B_2$, $B_3$, $B_4$, and $B_5$ sequentially. Results show that our method consistently outperforms the two baseline methods with lower average CERs on the OTs and the NTs in all stages, especially in stages 4 and 5 where the distribution and amount of NT data are more different from the OT data.}  
	\label{table3}  
	\resizebox{0.95\linewidth}{!}{
	\begin{tabular}{c||c||c|c||c|c|c||c|c|c|c||c|c|c|c|c}  
		\hline  
		& \bf Stage 1 & \multicolumn{2}{c||}{\bf Stage 2} & 
		\multicolumn{3}{c||}{\bf Stage 3} &
		\multicolumn{4}{c||}{\bf Stage 4} &
		\multicolumn{5}{c}{\bf Stage 5} \\
		\hline
		& Initial & OT & \multicolumn{1}{c||}{NT} & \multicolumn{2}{c|}{OT} & \multicolumn{1}{c||}{NT} & \multicolumn{3}{c|}{OT} & \multicolumn{1}{c||}{NT} & \multicolumn{4}{c|}{OT} & \multicolumn{1}{c}{NT}\\
		\hline
		& $B_1$ & $B_1$ & $B_2$ & $B_1$ & $B_2$ & $B_3$ & $B_1$ & $B_2$ & $B_3$ & $B_4$ & $B_1$ & $B_2$ & $B_3$ & $B_4$ & $B_5$ \\
		\hline
		Joint training&	5.10&	4.10&	2.09&	 4.27&	2.32&	1.78&	4.06&	2.48&	1.99&	6.30&	 4.07&	 2.48&	 1.98&	6.32&	4.06\\
		\hline
		\hline
		Fine-tuning	&5.10&	8.22&	2.14&	10.11&	13.24&	1.66&	28.10&	34.00&	28.77&	6.56&	27.36&	36.56&	28.87&	10.27&	3.89 \\
		\hline
		\hline
		RBKD \cite{ref12}&	5.10&	4.53&	\bf 2.17&	5.76&	5.21&	\bf 1.67&	12.44&	20.15&	13.19& 	\bf 6.73&	13.83&	20.32&	14.15&	10.06&	3.98 \\
		\hline
        RBKD+EWC \cite{ref12}&	5.10&	{\bf 4.12}&	2.58&	4.75&	2.79&	2.20&	6.22&	11.80&	7.62&	8.94&	6.47&	11.85&	7.90&	9.64&	4.68 \\
        \hline
        Our method&	5.10&	4.14&	2.18&	\bf 4.55&	\bf 2.72&	1.77&	\bf 5.31&	\bf 8.78&	\bf 5.41&	6.90&	\bf 5.10&	\bf 9.11&	\bf 6.28&	\bf 8.32&	\bf 3.97 \\
        \hline
	\end{tabular}
	}
\end{table*}

\subsection{Main experiment: multi-stage sequential training}
\textbf{A 5-stage sequential training scenario:} Consider the following 5 data subsets: $B_1$: AISHELL-1, $B_2$: Primewords, $B_3$: ST-CMDS, $B_4$: new accents data, $B_5$: new words data. In the 1\textsuperscript{st} stage, an ASR model is trained on the training set of $B_1$ based on the descriptions in Section 4.1. In the 2\textsuperscript{nd} stage, the model pre-trained for $B_1$ should be modified, given the training set of $B_2$, to perform well on both $B_1$ (the original task) and $B_2$ (the new task). The subsequent three stages are similar to the 2\textsuperscript{nd} stage, where in each stage, the model from the previous stage should be modified to adapt to a new task (i.e. $B_3$, $B_4$, and $B_5$, sequentially) given its associated training set.

\textbf{Performance evaluation protocol:} We use CER as the metric for performance evaluation. For stage $s\in\{2,3,4,5\}$, the learned model's performance on original tasks is jointly reflected by its CER on test set for $B_1, \cdots, B_{s-1}$; while its performance on the new task is reflected by the CER on the current test set for $B_s$.

\textbf{Methods and implementation:} We compare our method with the two incremental learning methods in \cite{ref12}: a vanilla RBKD method and a RBKD method combined with EWC -- to our knowledge, these are the only two existing incremental learning methods for ASR without access to original task data. We also compare our method with the fine-tuning method mentioned in Section 1. For all these methods, in each sequential training stage, {\it only} the training set for the new task is exploited. We also consider the joint training method, which, in each stage, exploits training data accumulated from all previous and current stages; thus, it is supposed to demonstrate a target-reference performance for both new tasks and original tasks \cite{ref30}. For our method, we set the hyperparameters as $T=3$, $\beta=0.03$ and $\gamma=500$, while their choices will not significantly influence the performance of our method, as will be shown in Section 4.3.

\begin{figure}[t]
	\centering
	\subfigure[Performance on the original tasks]{
		\begin{minipage}[t]{0.5\linewidth}
			\centering
			\includegraphics[width=\linewidth]{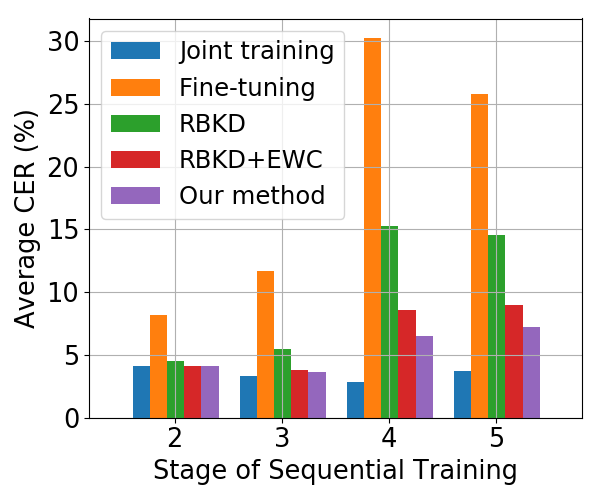}
		\end{minipage}
	}%
	\subfigure[Performance on the new tasks]{
		\begin{minipage}[t]{0.5\linewidth}
			\centering
			\includegraphics[width=\linewidth]{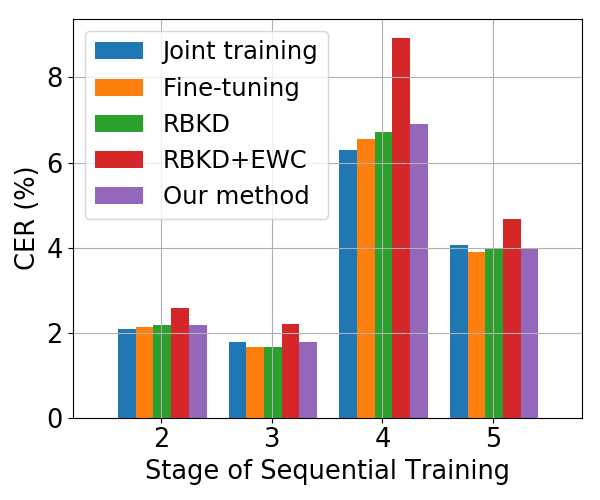}
		\end{minipage}
	}%
	\centering
	\caption{Average CERs on the original tasks and CERs on the new tasks for stage 2-5 of the 5-stage sequential training, for the 5 methods.}
	\label{fig3}
\end{figure}

\textbf{Results and discussion:} In Table \ref{table3}, for each of the 5 sequential training stages, we show CERs of the learned model on the test set for tasks ever visited, for the 5 methods. For better visual comparison between these methods, in Figure \ref{fig3}, for stage 2, 3, 4, and 5, we show the average CER for each method over all tasks ever visited, and also the CER for the task associated with the current stage. Clearly, fine-tuning suffers from catastrophic forgetting, as its average CERs on original tasks are uniformly large for all stages. RBKD alleviates the forgetting to some extent compared with fine-tuning, but its performance on original tasks is still not satisfactory. RBKD combined with EWC shows even lower average CERs than the vanilla RBKD (though still not comparable to our method) on original tasks; but its CERs on the new tasks are larger than other methods for all stages. This is consistent with our discussion in Section 2 that EWC may over-constrain the model parameters, which possibly leads to impaired learning for new tasks. In comparison, on the original tasks, our method clearly outperforms fine-tuning, RBKD, and RBKD with EWC; on the new tasks, our method shows similar performance as RBKD and fine-tuning, while outperforming RBKD with EWC clearly. In particular, the effectiveness of our novel EBKD mechanism in maintaining the model's performance on original tasks is clearly shown as our method achieves lower average CERs than the RBKD-based methods.

\subsection{Choice of hyperparameters}

In principle, when $T$ is large, predictions made by the pre-trained model with high confidence will be influential to our model training \cite{ref31}. Since the pre-trained model usually performs poorly (and also with low confidence) on new tasks, a large $T$ will likely block the ``wrong guidance'' from the pre-trained model regarding new tasks. For $\beta$ and $\gamma$, since they are weights for the terms devoted to maintaining the trained model's performance on original tasks, setting them overly large may prevent the model from learning for new tasks. Here, we study the choice of these three hyperparameters. For simplicity, we consider a single-stage incremental learning on the new accents data, with an ASR model pre-trained on the combined training set from the JD in-house data and the public data. Let $T=3$, $\beta=0.03$ and $\gamma=500$ (used in Section 4.2) be the baseline choice of the hyperparameters. Each time, we vary one hyperparameter with the other two fixed to the baseline choice for performance evaluation. Specifically, we consider $T\in\{1,2,3,4,5\}$, $\beta\in\{0.01,0.02,0.03,0.04,0.05\}$, and $\gamma\in\{ 100,200,500,1000,2000\}$, respectively. As shown in Figure \ref{fig2}, with the above-mentioned hyperparameter choices, the model trained by our method achieves a significant CER decreasing (2.7{\%}--3.8{\%}) on the new tasks, and only a small CER increment (less than 0.5{\%}) on the original tasks when compared with the pre-trained model. In general, within a proper range, the choice of the hyperparameters does not significantly affect the performance of our method. In practice, one can choose these hyperparameters using a small, held-out validation set.

\subsection{Results of two practical scenarios}

In practice, an ASR model pre-trained on a large data set for some tasks may need to be adapted to a new task. Here, we compare the three incremental learning methods (including ours), fine-tuning, and joint training (the target-reference method), on adapting a pre-trained ASR model to a new task on the new accents data (scenario $I$) and a new task on the new words data (scenario $J$), respectively. Model pre-training is performed on a combined training set from the JD in-house data and the public data. Table \ref{table4} shows the CER performance of the above-mentioned methods (and also the pre-trained model itself for reference) on the original tasks and the new tasks for both scenarios $I$ and $J$. Similar to the experiment in Section 4.2, for both (close-to-practice) scenarios, fine-tuning, RBKD, and RBKD with EWC more or less suffer from forgetting the original tasks. In comparison, the performance of our method on the original tasks is strong for both scenarios -- even comparable with the performance of joint training which directly uses the training set for both new tasks and original tasks. On the new tasks which the pre-trained model should be adapted to, our method clearly outperforms the two baseline methods, and is comparable with fine-tuning and joint training for the two scenarios in general. Numerically, compared to the target-reference joint training method, the performance drop (i.e. the average CER increment for all test sets in both scenarios) of our method is 0.02$\%$ CER, which is 97$\%$ smaller than the drops of the baseline methods.

\begin{table}[t]
	\centering  
	\caption{CERs of the three incremental learning methods (including ours), fine-tuning, and joint training on the Original Tasks (OTs) and New Tasks (NTs), for adapting a pre-trained ASR model to a new accent (scenario $I$) and new words (scenario $J$), respectively. The performance of the pre-trained model itself is included for reference.}  
	\label{table4}  
 	\resizebox{0.95\linewidth}{!}{
	\begin{tabular}{c|c|c|c|c|c}  
		\hline  
		& \multicolumn{2}{|c}{\textbf{CER on OTs}}
		& \multicolumn{2}{|c}{\textbf{CER on NTs}}
 		& \multicolumn{1}{|c}{\textbf{Avg. CER}}\\

		\hline
		& Sce. $I$ & Sce. $J$ & Sce. $I$ & Sce. $J$ & Sce. $I$ and $J$\\
		\hline
		\hline
        Pre-training & 4.46 & 4.46 & 10.07 & 18.47 &9.37\\
		\hline
		\hline
		Joint training & 4.79 & 4.79 & 6.70 & 1.63 & 4.48\\
        \hline
        \hline
        Fine-tuning & 6.02 & 7.69 & 7.19 & 1.49 & 5.60\\
        \hline
        \hline
        RBKD \cite{ref12} & 5.62 & 5.76 & 6.97 & 2.37 &5.18\\
        \hline
        RBKD+EWC \cite{ref12} & \bf 4.77 & 4.90 & 9.24 & 6.99 & 6.48\\
        \hline
        Our method & 4.78 & \bf 4.70 & \bf 6.82 & \bf 1.52 & \bf 4.46\\
        \hline
	\end{tabular}
 	}
\end{table}

\section{Conclusions}
\label{sec:Conclusions}

We propose an incremental learning method for end-to-end ASR models' adaptation to new speech recognition tasks without obviously degrading its performance on the original task it is trained for. To overcome forgetting, our method proposes a training loss containing a novel EBKD loss and a RBKD loss based on the pre-trained model. Our method does not require access to the original task data of the pre-trained model; and achieves the best CER performance on both original tasks and new tasks when compared with the baseline methods.



\end{document}